\documentclass[preprint,12pt, 3p, titlepage]{elsarticle} 
\usepackage{graphicx}
\usepackage{latexsym}
\usepackage[utf8]{inputenc}
\usepackage{amsthm}
\usepackage{mathtools}  
\usepackage{amsmath}
\usepackage{amssymb}
\usepackage{amsfonts}
\usepackage{multirow}
\usepackage{caption,subcaption}
\usepackage{url}
\usepackage{float}
\usepackage{xcolor}
\usepackage{booktabs}
\usepackage{adjustbox}
\usepackage{soul}
\usepackage{lineno}
\usepackage{caption,subcaption}
\usepackage{setspace}
\usepackage{fixmath}
\usepackage{adjustbox}
\usepackage{array} 

\usepackage{lineno}

\allowdisplaybreaks

\usepackage[ruled,algonl,vlined]{algorithm2e}
\usepackage{fixltx2e}

\usepackage{lineno}

\usepackage{lineno}

\newcommand{\vikasR}[1]{{#1}}

\usepackage{natbib}
\setcitestyle{round, comma, authoryear}

\begin{document}

\begin{frontmatter}
\title{Data augmentation and refinement for recommender system: A semi-supervised approach using maximum margin matrix factorization}

    \author[1]{Shamal Shaikh}
    \ead{ss911876@student.nitw.ac.in}
    \author[1]{Venkateswara Rao Kagita}
    \ead{venkat.kagita@nitw.ac.in}
    \author[2]{Vikas Kumar\corref{cor1}}
    \ead{vikas007bca@gmail.com}
    \author[3]{Arun K Pujari}
    \ead{arun.pujari@mahindrauniversity.edu.in}
    \address[1]{National Institute of Technology, Warangal, India}
    \address[2]{University of Delhi, New Delhi, India}
    \address[3]{Mahindra University, Hyderabad, India} 
    
    \cortext[cor1]{Corresponding author}

\begin{abstract}

Collaborative filtering (CF) has become a popular method for developing recommender systems (RSs) where ratings of a user for new items are predicted based on her past preferences and available preference information of other users. Despite the popularity of CF-based methods, their performance is often greatly limited by the sparsity of observed entries. In this study, we explore the data augmentation and refinement aspects of Maximum Margin Matrix Factorization (MMMF), a widely accepted CF technique for rating predictions, which has not been investigated before. We exploit the inherent characteristics of CF algorithms to assess the confidence level of individual ratings and propose a semi-supervised approach for rating augmentation based on self-training. We hypothesize that any CF algorithm's predictions with low confidence are due to some deficiency in the training data and hence, the performance of the algorithm can be improved by adopting a systematic data augmentation strategy. We iteratively use some of the ratings predicted with high confidence to augment the training data and remove low-confidence entries through a refinement process. By repeating this process, the system learns to improve prediction accuracy. Our method is experimentally evaluated on several state-of-the-art CF algorithms and leads to informative rating augmentation, improving the performance of the baseline approaches.

\end{abstract}


\end{frontmatter}

\section{Introduction}
    A recommender system (RS) is a subset of information filtering systems designed to deal with information overload. These systems sift the users through the information space and eliminate the need for manual filtering of their possible choices over the entire product space. Recently, RS has been widely and successfully applied in several different applications,  including e-commerce~\citep{hidasi2015session}, e-library~\citep{wang2018content}, e-learning~\citep{manouselis2011recommender}, tourism~\citep{wang2017your}, job recommendation ~\citep{dave2018combined}, and drug recommendation~\citep{he2018kernelized}. RS attempts to analyze the feedback of the users and recommends a product/item to a user based on past feedback of the users about the product/item and the user’s taste based on past purchase history. Any RS aims to construct a probable recommendation set of items tailored for individual users that matches their needs and taste. In the literature, several techniques have been proposed to generate a recommendation that can be categorized into content-based (CB), collaborative filtering (CF), and hybrid approaches~\citep{bobadilla2013recommender,burke2002hybrid}. The CB approach generates the recommendation based on the match between the user’s profiles and the item’s contents. In contrast, the CF approach uses users’ past preferences for items that are available in the form of feedback, either explicit, such as ratings and reviews, or implicit, discovered based on the actions that the user performs concerning items. On the other hand, the hybrid approach combines different techniques of CF and CB approaches. 
    
    Among several approaches for Recommender systems (RSs), collaborative filtering (CF) has emerged as a fundamental paradigm~\citep{guo2014merging, adomavicius2005toward, mobasher2003semantically}.  Despite the popularity of CF-based methods, their performance is often greatly limited by the number of observed entries.  To alleviate the problem of data sparsity, researchers have introduced semi-supervised learning. Semi-supervised learning is a crucial concept that focuses on exploiting knowledge from a large amount of unlabeled data and the small amount of available labeled data. Most existing CF methods adopt domain adaptation wherein knowledge learned from side information such as review texts and pictural data like posters~\citep{yu2020semi, zhao2016matrix} is transferred to the main domain. However, exploiting and integrating side information requires consistent knowledge across auxiliary domains~\citep{zhang2020cross, lu2013selective}. For example, the knowledge captured from book reviews is consistent with the knowledge in the movie domain, as both domains overlap in genre representations. However, it is inconsistent with the knowledge in the clothing domain~\citep{liu2019jscn}.   Thus, there is a need to implement a solution that can handle data sparsity without the involvement of side information or identification of compatible domains.

    In this paper, we propose a self-training-based CF approach for rating augmentation that eases the burden of finding compatible domains for knowledge acquisition. In machine learning research, many self-training based have been proposed for the classification task, and the area of recommender system has been largely untouched with self-training based models~\citep{gao2022enhancing, wu2018self, nartey2019semi, xie2020self}. The idea is to leverage the learning process by utilizing both a small portion of label data and a large amount of unlabelled data~\citep{krishnapuram2004semi}. The recommendation task over ordinal rating preferences can also be visualized as a classification task~\citep{rennie2005fast, kumar2017proximal, kumar2017collaborative}. 
    In~\citep{kumar2017collaborative}, maximum margin matrix factorization (MMMF), a popular recommendation model that provides prediction with high accuracy, is visualized as an extension of a two-class classifier to a unified multi-class classifier and proposed a method for ordinal matrix completion by hierarchically arranging multiple binary MMMF.   
    This paper advances the frontier of research on this subject by investigating the data augmentation and refinement aspects which are not attempted so far by the researchers. The inherent characteristics of MMMF are exploited to assess the confidence level of algorithm’s prediction of individual rating. We hypothesize that the any CF algorithm’s predictions with low confidence can be due to some deficiency in the training data and hence, the performance of the algorithm can be improved by adopting a systematic data augmentation strategy. In the present study, we consider Maximum Margin Matrix Factorization as the base CF-based algorithm. We propose a self-training-based semi-supervised approach for rating augmentation to improve the performance of MMMF. The generic principle that we adopt here is to assess the algorithm’s prediction confidence qualitatively as low or high. Our method is an iterative process in which the ratings of unknown items predicted with high confidence are used to augment the training data. At this stage, the refinement process removes an entry in the training data that is found to be of low confidence. By repeated application of the process of augmentation and refinement, the system learns to improve prediction accuracy. We experimentally evaluate the proposed rating augmentation technique by considering several state-of-the CF algorithms. The experimental result corroborates our claims that the proposed strategy leads to informative rating augmentation and improves the performance of the baseline approaches under consideration.

    The rest of the paper is organized as follows.  Section~\ref{sec:RW} briefly reviews existing approaches to handle data sparsity issues in the recommendation system. Section~\ref{sec:mmmf} summarizes the well-known existing MMMF process. We present the proposed ST-MMMF model of rating augmentation in Section~\ref{PS}.  An experimental analysis of the proposed algorithm is reported in Section~\ref{ES}. Finally, Section~\ref{CFW} concludes and indicates several issues for future work.

\section{Related Work} 
\label{sec:RW}
 The traditional approaches to alleviating the problem of data sparsity
 can be broadly divided into two categories - methods that impute ratings to generate pseudo-rating data based on some estimation criterion and methods that use auxiliary information~\citep{kuo2021application, natarajan2020resolving, duan2022combining}. In this section, we present a brief overview of the significant approaches proposed to overcome the problem of data sparsity in the user-item rating matrix. 

Rating imputation is a non-trivial task and requires sophisticated methods to infer the missing rating. Several approaches have been proposed in recent years for rating imputation, among which the mean-based imputation is the most naive. The rating for any unobserved (user, item)-pair can be imputed either with the user-specific mean or item-specific mean~\citep{breese2013empirical, ghazanfar2013advantage}. The mean method of rating imputation suffers from high bias because there are going to be so many entries in the rating matrix with a similar rating level which may lead to an imbalance rating distribution. Matrix factorization techniques, such as Singular Value Decomposition (SVD) and Non-negative Matrix Factorization (NMF), have been widely used for imputing missing ratings in recommender systems~\citep{KIM2005823, ranjbar2015imputation}. These methods factorize the rating matrix into two low-rank matrices and use the factorized matrices to estimate the missing ratings.  In~\citep{hwang2016told}, SVD is applied on the pre-use preferences matrix to identify the uninteresting items and then assigns zero ratings. Collaborative filtering methods, such as user- and item-based, have also been used to rating imputation~\citep{bell2007modeling, ghazanfar2013advantage}.  These methods use the similarity between users or items to estimate missing ratings~\citep{ren2012efficient}. Bayesian methods, such as Probabilistic Matrix Factorization (PMF) and Bayesian Personalized Ranking (BPR), have been used to incorporate uncertainty in the imputation process~\citep{mnih2007probabilistic, rendle2012bpr, wang2019enhancing, moon2023comix}. These methods model the rating matrix as a probabilistic generative model and use Bayesian inference to estimate missing ratings.  The other category of methods uses side information, such as reviews, images, and videos, to handle the data sparsity problem~\citep{guo2019exploiting}. Niu et al.~\citep{niu2016fuir} proposed a neighborhood-based approach of collaborative filtering where the reviews are used as side information to compute the similarity between two users.  Ning et al.~\citep{ning2012sparse} propose several methods to utilize the item side information to learn sparse linear coefficient matrix to the top-N recommendation. In~\citep{strub2016hybrid}, a Neural Network architecture called CFN  to learn the non-linear representation of users and items. The authors introduce a novel loss function adapted to input data with missing values and utilize the side information. A deep hybrid model is proposed to learn deep users' and items’ latent factors from the rating matrix and side information~\citep{dong2017hybrid}. In~\citep{massa2004trust}, a trust matrix and the user-item rating matrix are used together to improve the accuracy of similarity calculation and recommendation quality.

\section{Maximum Margin Matrix Factorization}
\label{sec:mmmf}
Maximum Margin Matrix Factorization (MMMF) is a well-known method for computing a dense approximation $X \in \mathbb{R}^{N \times M}$ of a sparse matrix $Y \in \mathbb{R}^{N \times M}$ with ordinal entries~\citep{srebro2004maximum}. The initial proposal for MMMF is formulated as a semi-definite programming (SDP) problem that can handle the factorization of small matrices. Rennie et al.~\citep{rennie2005fast} proposed a gradient-based optimization method for MMMF to factorize a matrix with sufficiently large size. We will refer to this as MMMF in our subsequent discussion. The readers are requested to refer~\citep{rennie2005fast, kumar2019collaborative} for details. 

\subsection{MMMF as Gradient-based Optimization}

The MMMF approach is primarily designed for collaborative filtering tasks where the user's preferences over items are observed in the form of ratings and are naturally organized in matrix form. Let $Y  = [y_{ij}] \in \mathbb{R}^{N \times M}$ denote the rating matrix, where $N$ represents the number of users, and $M$ is the number of items. The entries $y_{ij}$ are from $\{0, 1, 2, \dots, R\}$, where R denotes the maximum rating level, and $0$ denotes the unobserved entries.  Given a sparse rating matrix $Y$, MMMF seeks a minimum trace norm matrix  $X \in \mathbb{R}^{N \times M}$ that approximates the observed entries in matrix $Y$. The gradient-based optimization problem with a term $\|X\|_{\Sigma}$,  trace norm of matrix $X$, is a complicated non-differentiable function for which it is difficult to find the subdifferential~\citep{rennie2005fast}. Hence, instead of searching over $X$, MMMF learns a pair of low norm factor matrices $U\in \mathbb{R}^{N \times d}$ and $V\in \mathbb{R}^{M \times d}$  such that

\begin{equation}
Y \approx UV^T,
\end{equation}

\noindent where $d$ is a parameter. 

The factor matrices $U$ and $V$ are obtained by minimizing the regularized loss function where the Froebenius norm of the factor matrices is used to upper bound the minimization objective with the term $\|X\|_{\Sigma}$, i.e., for any $U$ and $V$, $\|UV^T\|_{\Sigma} \le \frac{1}{2}(\|U\|_{F}^2 + \|V\|_{F}^2)$. In order to map the predicted value $x_{ij} = U_{i} V_{j}^T$ into $R$ intervals, MMMF also learns $R-1$ thresholds, $\theta_{i,1} \le \theta_{i,2}, \dots, \le \theta_{i,R-1} $, for each $i$th user. The objective function of MMMF for ordinal
rating predictions can be written as follows.

\begin{equation}
\label{eq:mmmfFormulation1}
\underset{U, V}{min}\;J(U,V,\Theta) = \sum_{y_{ij}|ij \in \Omega}{ \big( \sum_{r=1}^{y_{ij} - 1} {h(U_iV_j^T - \theta_{i,r})} + \sum_{r=y_{ij}}^{R-1} {h(\theta_{i,r} - U_iV_j^T)} \big) }  + \frac{\lambda}{2}(\|U\|^{2}_{F} + \|V\|^{2}_{F}),
\end{equation}

\noindent 
where $\|\cdot\|_F$ is Frobenius norm, $U_i$ and $V_j$ denotes the $i$th row of matrix $U$ and matrix $V$, respectively,  $\Theta = [\theta_{ij}]$ is a threshold matrix consisting of $R-1$ thresholds for each $i$th user, $\lambda > 0$ is  the regularization parameter, $\Omega$ is the set of observed entries and $h(z)$ is the smooth hinge loss  defined as

\begin{equation}
 h(z) = 
  \begin{cases}
  0, & \text{if z $\geq$ 1}; \\
  \frac{1}{2}(1 - z)^{2}, & \text{if $0 < z < 1$};\\
  \frac{1}{2} - z, & \text{otherwise}.
  \end{cases}
  \label{smoothHinge} \\
\end{equation}

\noindent The equation given in (\ref{eq:mmmfFormulation1}) can be rewritten as follows.
\begin{equation}
\label{mmmfOptimizationProblem}
\underset{U, V}{min}\;J(U,V,\Theta) = \sum_{r = 1}^{R-1}{\sum_{(i,j) \in \Omega}h \big(T_{ij}^{r}(\theta_{i,r} - U_{i}V_{j}^{T}))} + \frac{\lambda}{2}(\|U\|^{2}_{F} + \|V\|^{2}_{F})
\end{equation}

\noindent
where $T$ is defined as

\begin{equation*}
T_{ij}^r = 
\begin{cases}
+1, & \text{if r $\geq$ $y_{ij}$;} \\
-1, & \text{if r $ < $ $y_{ij}$.}
\end{cases}
\end{equation*}

\noindent Several approaches can be used to optimize the objective function \ref{mmmfOptimizationProblem}. A gradient descent method and its variants start with random $U$, $V$ and $\Theta$ and iteratively update them using equations \ref{U_update}, \ref{V_update} and \ref{theta_update}, respectively.
\begin{align}
 U_{ip}^{t+1} &= U_{ip}^{t} - c \frac{\partial J}{\partial U_{ip}^t} \label{U_update}.\\
 V_{jq}^{t+1} &= V_{jq}^{t} - c \frac{\partial J}{\partial V_{jq}^t} \label{V_update}.\\
 \theta_{i,r}^{t+1} &= \theta_{i,r}^{t} - c \frac{\partial J}{\partial \theta_{i,r}^t} \label{theta_update}.\
\end{align}
$c$ is the learning rate in the above equations and suffixes $t$ and $(t + 1)$ indicate current and updated values. The gradients of the variables to be optimized are determined as follows. 

\begin{align}
\frac{\partial J}{\partial U_{ip}} &= \lambda U_{ip} - \sum_{r = 1}^{R-1}{ \sum_{j|(i,j) \in \Omega}{ T_{ij}^{r} h' \big(T_{ij}^{r}(\theta_{i,r} - U_{i}V_{j}^{T})\big)V_{jp}  } } \\
\frac{\partial J}{\partial V_{jq}}  &= \lambda V_{jq} - \sum_{r = 1}^{R-1}{ \sum_{i|(i,j) \in \Omega}{ T_{ij}^{r} h' \big(T_{ij}^{r}(\theta_{i,r} - U_{i}V_{j}^{T})\big)U_{iq}  } } \\ 
\frac{\partial J}{\partial \theta_{i,r}} &= \sum_{j|(i,j) \in \Omega}{ T_{ij}^{r} h' \big(T_{ij}^{r}(\theta_{i,r} - U_{i}V_{j}^{T})\big)} 
\end{align}

\noindent
where $h'(z)$ is defined as follows.
\begin{equation} 
  h'(z) = 
  \begin{cases}
  0, & \text{if z $\geq$ 1}; \\
  z-1, & \text{if $0 < z < 1$};\\
  -1, & \text{otherwise}.
  \end{cases}
  \label{smoothHingederivation}
\end{equation}

\noindent Once $U$, $V$ and $\theta$'s are computed, the matrix completion process is accomplished as follows.
\begin{equation*}
\hat{y}_{ij} = 
\begin{cases}
r, & \text{if $(i,j) \notin \Omega \wedge (\theta_{i,r} \le x_{ij} \le \theta_{i,r+1}) \wedge (0 \le r \le R-1)$}; \\
y_{ij}, & \text{if $(i,j) \in \Omega$},
\end{cases}
\end{equation*}
where, $\hat{y}_{ij}$ is the prediction for item $j$ by user $i$. For simplicity of notation, we assume $\theta_{i,0} = -\infty$ and $\theta_{i,R} = +\infty$ for each user $i$.

\subsection{Geometrical Interpretation of MMMF}
\label{sec:geoMMMF}
\begin{figure} [ht!]
    \centering
    \adjustbox{max width=\linewidth}{
	\includegraphics[width=\linewidth,height=3.1in]{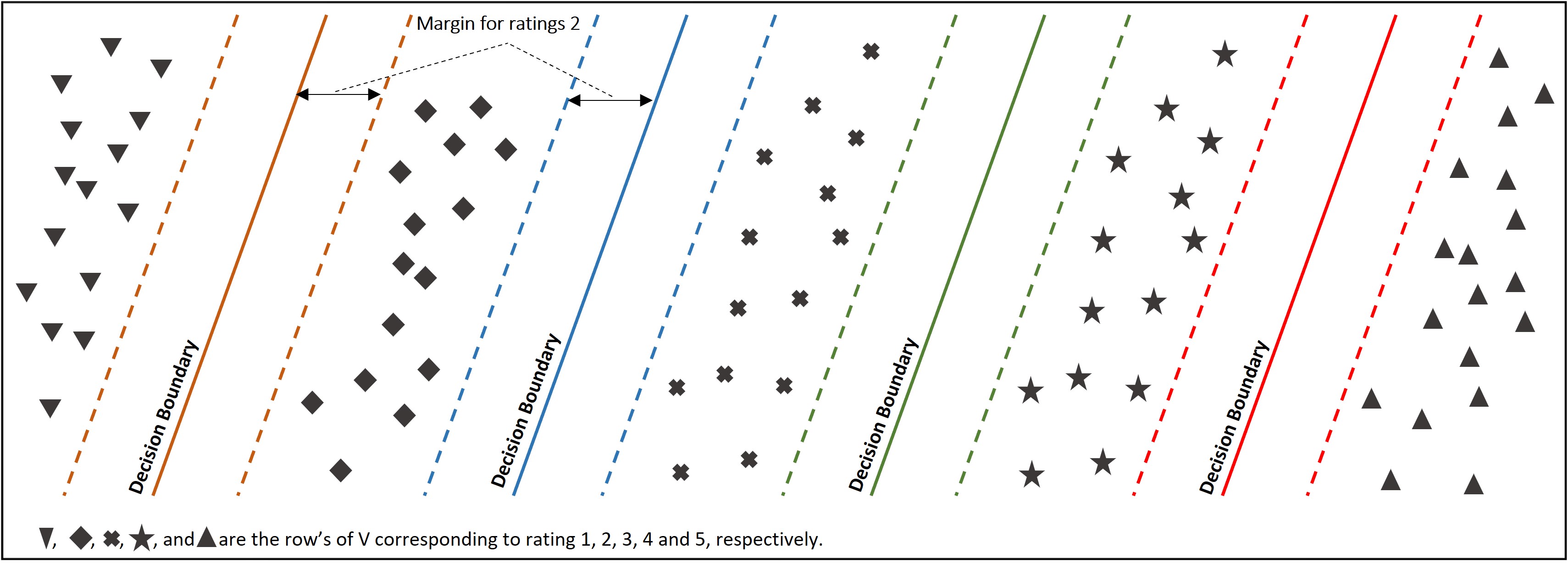}
    }
    \caption{Classification by MMMF for the ith user}
    \label{fig:geoMMMF}
\end{figure}

MMMF decomposes a large user-item rating matrix $Y$ into two low-norm matrices $U$ and $V$, the former representing users and the latter representing items. Geometrically, each row of user latent factor matrix $U$ defines a hyperplane in the $d$-dimensional space, and similarly, the rows of $V$ can be viewed as points in the same space. The objective is to learn from a sparse $Y$ the decision hyperplane for each user to separate points of diferent ratings with the largest possible margin. When $Y$ is bi-valued ($\{-1, +1\}$), the objective is to learn the hyperplane defining for each user ($U_i$ for user $i$) a decision function  that separates one rating ($+1$) from the other ($-1$) with maximum margin. In the case of ordinal scale, i.e., $y_{ij} \in \{1, 2, \dots R\}$, The objective is to partition the set of items embedded as points into R-regions, each corresponding to a rating level, separated by parallal hyperplanes defined by $U_i$ for user $i$.  This is achieved by optimizing a smooth version of the hinge loss function well-known for margin maximization~\citep{rennie2005fast, kumar2017collaborative,  kumar2017proximal, salman2016combining}. The outcome of the optimization is to get the optimal combination of $(U, V, \Theta)$ such that the embedding of items rated as $r$ fall as correctly as possible into regions defined by $(U_i, \theta_{i,r-1})$ and $(U_i, \theta_{i,r})$  with sufficient margin. For the simplicity of notation, we assume $\theta_{i,0} = -\infty$ and $\theta_{i,R} = +\infty$, for each $i$th user.  Figure~\ref{fig:geoMMMF} illustrate this concept by taking a row of $U$ as the decision hyperplane for a user, and rows of $V$ are embedding of points corresponding to the items.

\section{Proposed Algorithm}\label{PS}
In this section, we describe the proposed data augmentation and refinement strategy, denoted as Self-Training with Maximum Margin Matrix Factorization (ST-MMMF), which is trained in a semi-supervised fashion to generate new rating samples. Data augmentation in ML allows artificially increasing the size of the training set by adding new synthetic examples and helps the decision function to become invariant to the changes~\citep{xie2020unsupervised, ratner2017learning}. This can be interpreted as a regularization method that induces a useful bias by preventing the model from focusing on irrelevant features and making it less prone to overfitting ~\citep{chen2020group}. \textit{Data augmentation} in the present context implies that some synthetic user-item ratings are introduced where such data is not available in the original user-item interaction record. In other words, it is to introduce some values when $y_{ij}$ is unknown in $Y$. The choice of  $y_{ij}$ cannot be arbitrary, and hence, data augmentation is the judicious process of selecting unknown $y_{ij}$'s, which can be used to augment the sparse matrix $Y$.  We show later that our augmentation process satisfies the desirable properties required for the purpose.  The process of \textit{data refinement} is yet another strategy adopted in ML research to improve the system's performance. In the present context, we design the proposed data refinement strategy by identifying certain weak $y_{ij}$'s known in the original training $Y$ and remove them to refine the training set.

\vikasR{ST-MMMF follows an iterative process for \textit{data augmentation} and \textit{data refinement}.  Each iteration of ST-MMMF is comprised of four pivotal stages, (1) Learning the latent factor matrices $U$,  $V$, and $\Theta$ for the current $Y$, where $U~ (V)$ represents the user (item) latent factor matrix, and $\Theta$ signifies the threshold matrix; (2) Augmenting $Y$ with unobserved entries predicted with high confidence in Stage (1); (3) Refining the augmented $Y$ by removing the known ratings indicated in Stage (1) as of low confidence; (4) Replacing the matrix $Y$ with the refined and augmented matrix obtained after Stage (3) for the subsequent training iteration.}

\begin{figure}[ht!]
    \adjustbox{max width=1\linewidth}{
        \includegraphics[width=\linewidth, height=4.5in]{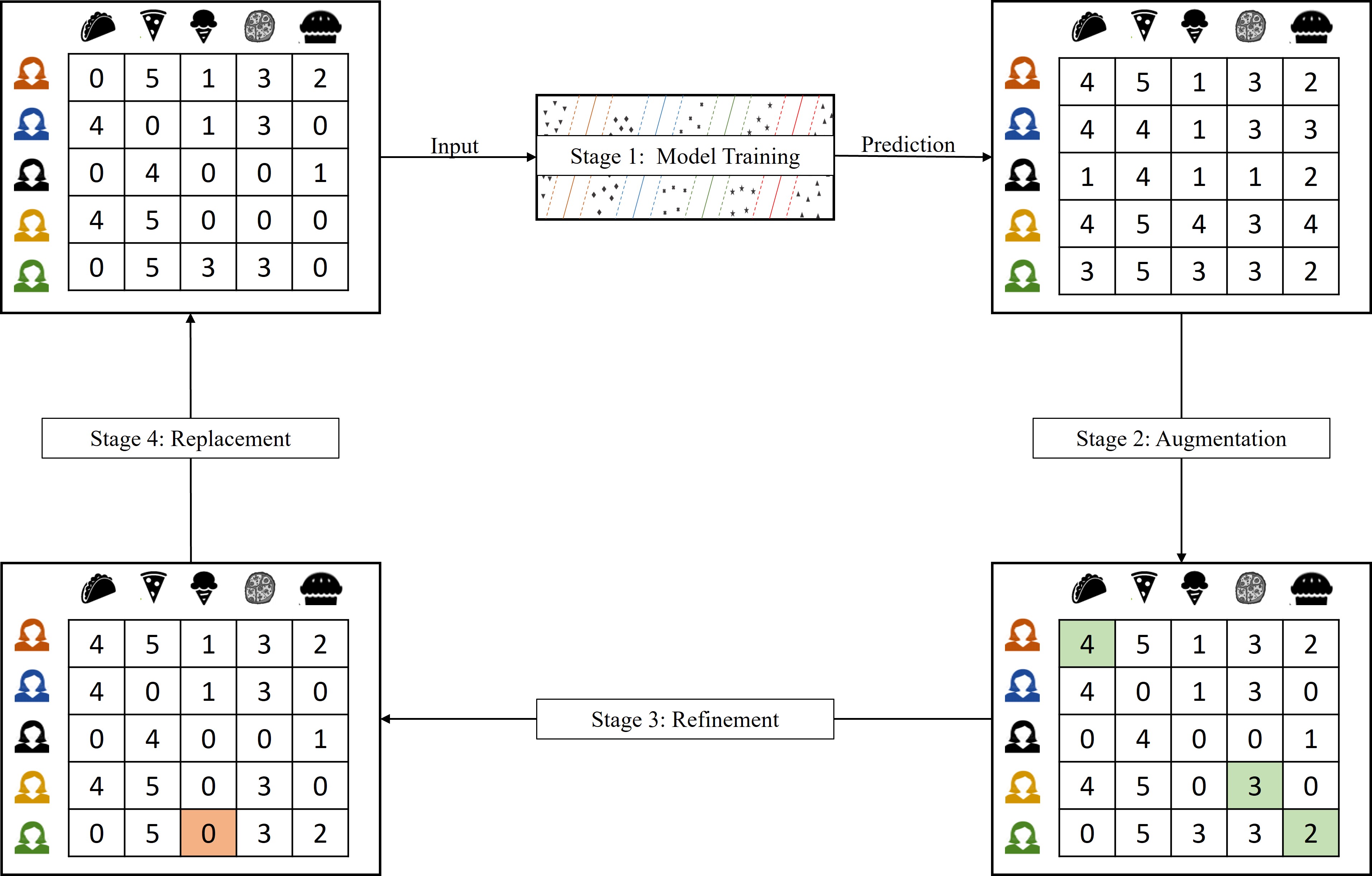}
    }
    \caption{Overview of the $i$th iteration of ST-MMMF.}
    \label{fig:ST-MMMF}
\end{figure}

 \noindent \vikasR{Figure~\ref{fig:ST-MMMF} depicts a high-level overview of the proposed ST-MMMF approach showing various stages involved in an $i$th iteration of the proposed ST-MMMF. The matrix at the top left corner denotes the input matrix $Y$ to $i$th iteration. In Stage 1, MMMF is employed to learn two latent factor matrices, $U$ and $V$, representing users and items. Additionally, $R-1$ thresholds are also learned for each user to map the corresponding real-valued predictions to the discrete rating scale. The matrix at the top right corner is the predicted matrix obtained after the application of the MMMF algorithm over the input matrix $Y$. Following the geometrical interpretation of MMMF, we have identified the highly confident predictions corresponding to the unobserved entries in the input matrix $Y$. Similarly, we have also marked the observed entries predicted with low confidence. The detailed procedure is explained in Section 4.1. In Stage 2, a subset of ratings identified to be predicted with high confidence is included in the matrix $Y$. The matrix at the bottom right corner shows the resultant matrix after the augmentation phase, with color-coded entries denoting the newly added ratings. Finally, in the refinement phase, the observed entries of matrix $Y$ marked to be predicted with low confidence in Stage 1 are removed. The matrix shown in the bottom left corner, obtained after Stage 3, is given as input to the next iteration of ST-MMMF.}

\subsection{Data Augmentation and Refinement}

In the original proposal of MMMF, the authors used the all-threshold hinge function to calculate the prediction loss corresponding to the observed entries in the rating matrix.  Geometrically, the all-threshold hinge function not only tries to embed the points rated as $r$ by the $i$th user into the region defined by $(U_i, \theta_{i, r-1})$ and $(U_i, \theta_{i,r})$ but also consider the position of the points with respect to other hyperplanes. This is more reasonable in discrete ordinal rating prediction, where it is always better to predict `2' than `5' if the true label is `1', unlike the multi-class classification setting where all mistakes are equal.  Hence, it is desirable that $V_j$, an embedding for $j$th item rated by the $i$th user, should satisfy the condition $U_iV_j^T - \theta_{i, r-1} > 0$ for $r < y_{ij}$ and $U_iV_j^T - \theta_{i,r} < 0$ for $r \ge y_{ij}$. For every $i$th user, the points falling in the region defined by $(U_i, \theta_{i, r-1})$ and $(U_i, \theta_{i, r})$  are assigned rating $r$ in the prediction. 

\begin{figure}[ht!]
    \adjustbox{max width=\linewidth}{
	\includegraphics[width=\linewidth, height=3in]{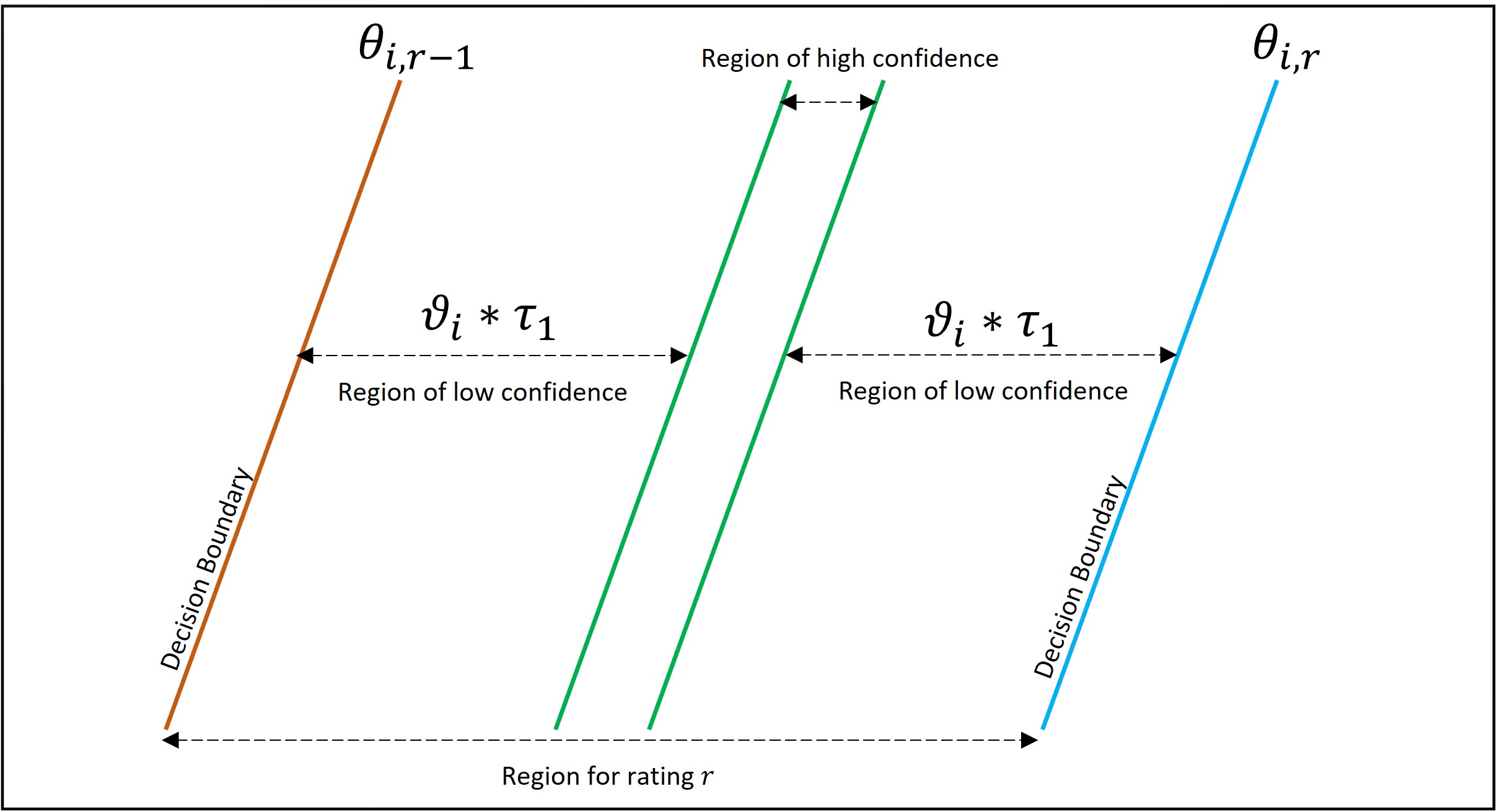}
    }
	\caption{Overview of ST-MMMF for rating imputation.}
	\label{fig:ST-MMMF_augRef}
\end{figure}

\begin{algorithm}[ht!]
    \setstretch{1.2}
    \SetAlgoLined
    \SetKwData{Left}{left}\SetKwData{This}{this}\SetKwData{Up}{up}
    \SetKwFunction{Union}{Union}\SetKwFunction{FindCompress}{FindCompress}
    \SetKwInOut{Input}{Input}\SetKwInOut{Output}{Output}
    \Input{Rating Matrix: $Y$, Size of Latent Dimension Space: $d$, Regularization Parameter: $\lambda_1$, Parameters for Rating Augmentation and Refinement: $\tau_1$ and $\tau_2$, Sampling Percentage = $s$}
    \Output{Predicted Rating Matrix: $\hat{Y}$}
    \BlankLine
    \Repeat{stop criterion reached}
    {
        $\Omega \leftarrow Index(Y \neq 0) $\\
        $Y^A = [0]^{N \times M}$ \\
        $U, V, \Theta \leftarrow MMMF(Y, d, \lambda)$ /* $U \in \mathbb{R}^{N \times d}, V \in \mathbb{R}^{M \times d}$ and $\Theta = [\theta_{ij}] \in \mathbb{R}^{N \times R-1} $ */ \\   
        $X \leftarrow UV^T$\\
        $\hat{Y} \leftarrow$ Discretize ($X$, $\Theta$) \\
        \For{$i\leftarrow 1$ \KwTo $N$}
        {
            $\vartheta_i \leftarrow$ AvgDist($\theta_{i,1}, \theta_{i,2}, \dots, \theta_{i,R-1} $) \\
        }
        /* Augmentation */\\
        \ForEach{$ij \notin \Omega $}
        {%
            \For{$r\leftarrow 1$ \KwTo $R$}
            {
                     /* $\theta_{i,0} \leftarrow -\infty$ and $\theta_{i,R} \leftarrow +\infty $ */\\
                \lIf{find `r' satisfying Equation~(\ref{eq:augment})}
                {
                    $y^A_{ij} \leftarrow r$                
                }
            }
        }

        /* Refinement */\\
        \ForEach{$ij \in \Omega $}
        {
            \For{$r\leftarrow 1$ \KwTo $R$}
            {
                \lIf{ find `r' satisfying Equation~(\ref{eq:refine})}
                {
                    $y_{ij} \leftarrow 0$                
                }
            }
        }    
        $Y \leftarrow Y +$ Sample($Y^A,s$) 
    }
    \caption{ST-MMMF ( $Y$, $d$, $K$, $\lambda$, $\tau_1$, $\tau_2$)}
    \label{algo:ST-MMMF}
\end{algorithm} 

ST-MMMF data augmentation and refinement process exploits the geometrical interpretation of MMMF. We assume that the region towards the center defined by $(U_i, \theta_{i, r-1})$ and $(U_i, \theta_{i,r})$ contains the points for which rating $r$ is predicted with high confidence. Hence, the unobserved entry $y_{ij}$ satisfying the following condition is augmented with rating $r$ for the next round of the training process. 

\begin{equation}
\label{eq:augment}
     \theta_{i,r-1} + \vartheta_i * \tau_1 < U_iV_j^T <  \theta_{i,r} - \vartheta_i * \tau_1
\end{equation}

\noindent Here, $\vartheta_i = \frac{1}{R-2}\sum_{r=2}^{R-1}{(\theta_{i,r} - \theta_{i,r-1}) }$ is the average gap between consecutive pairs of threshold values learnt for the $i$th user. $\tau_1 < 0.5$ is a shifting parameter used to define the region towards the center bounded by $(U_i, \theta_{i, r-1})$ and $(U_i, \theta_{i,r})$. In other words, Equation~(\ref{eq:augment}) defines the region of high confidence for the $i$th user. Any unobserved entry $y_{ij}$ satisfying Equation~(\ref{eq:augment}) is then correctly classified as rating $r$ with high confidence. The high-confidence region for the rating augmentation is depicted in Figure~\ref{fig:ST-MMMF_augRef}. \noindent Similarly, the known $y_{ij}$ falling  closer the boundary  defined by $(U_i, \theta_{i,r})$ can be termed as input data with low confidence. Hence, the observed entry $y_{ij}$ satisfying the following condition is discarded for the next round of the training process.

\begin{equation}
\label{eq:refine}
     \theta_{i,r} - \vartheta_i * \tau_2 < U_iV_j^T <  \theta_{i,r} + \vartheta_i * \tau_2
\end{equation}
Here, $\vartheta_i$ is the same as defined previously, and $\tau_2 < \tau_1$ is a shifting parameter. Algorithm~\ref{algo:ST-MMMF} outlines the main flow of the proposed method where the training set is augmented by adding all the ratings selected with $(\tau_1,s)$. 

\subsection{Handling Skewed Dataset}\label{PS1}

The ratings are nonuniformly distributed in many real-world datasets and therefore, in the data augmentation process, the rating label with the highest proportion tends to be augmented more frequently than that with a smaller proportion. This will result in the eventual degradation of the quality of the training set. We take into account rating distribution in the current training set while performing the augmentation to overcome the possible degradation. The frequency of augmenting rating labels with low distribution is made more frequently than those with high distribution.  Let $\mathbb{Z}_i$  denote the ratio of $i$th label rating in the training set, and $ S (\tau_1,s)$ is the set of ratings selected with the hyperparameters $(\tau_1,s)$ for the augmentation. We compute the total number of samples to be augmented to each rating label from the set $ S (\tau_1,s)$ on the following basis.
\begin{equation}
    \frac{ (1- \mathbb{Z}_1)}{\sum_{j = 1}^{R} (1-\mathbb{Z}_j)} : \frac{ (1- \mathbb{Z}_2)}{\sum_{j = 1}^{R} (1-\mathbb{Z}_j)}:\dots:\frac{ (1- \mathbb{Z}_R)}{\sum_{j = 1}^{R} (1-\mathbb{Z}_j)}
\end{equation}

\noindent Furthermore, to avoid the drastic change in rating distribution, the training set can be augmented with a small proportion of ratings predicted with high confidence.   The maximum number of ratings that can be augmented in each iteration of ST-MMMF is set to $5000$ for our experimental analysis.

\subsection{Complexity Analysis}
\vikasR{
In this section, we analyze the computational complexity of the proposed method. The time complexity of ST-MMMF mainly comprises of two components: 1) Learning the latent factor matrices $U$ and $V$ corresponding to users and items, respectively, and a threshold matrix $\Theta$ consisting of $R - 1$ thresholds for each user; and 2) Computation requires for augmentation and refinement. The optimization problem of MMMF given in Equation~\ref{eq:mmmfFormulation1} requires major computations for matrix multiplication. For the simplicity of representation, we assume that the cost of multiplying two matrices of size $N \times d$ and $M \times d$ is $O(NMd)$. The updation of matrix $U$ requires $(R-1)$ time multiplication of two matrices of size  $N \times M$ and $M \times d$ in the worst case. A similar computation is required for the updation of matrices $V$ and $\Theta$. Hence, the overall computation required to learn the matrices $U$, $V$, and $\Theta$ using MMMF is  $O( 3t_1(R-1)(NMd))$, that is, $O(t_1RNMd)$, where $t_1$ is the maximum number of gradient iteration~\citep{veeramachaneni2019maximum}. The augmentation and refinement process inherently calculates two thresholds for each rating level defining the left and right boundary. During the augmentation phase, only unobserved entries are considered whereas the refinement phase takes into account observed entries. In summary, each real-valued prediction is compared with $(2 \times R)$ thresholds. Hence, the computation cost required for the augmentation and refinement phase is $O(2RMN)$, that is, $O(RNM)$. Let $t_2$ denote the maximum number of iterations required for ST-MMMF. Hence, the overall computation cost of the proposed method is $O(t_2(t_1RNMd + RNM))$, that is, $O(t_1t_2RNMd)$.
}

\section{Experiments}\label{ES}
This section reports the experimental analysis of the proposed ST-MMMF algorithm. The performance of the rating imputation task is evaluated in terms of the predictive performance of the proposed approach as compared to several existing baseline approaches.

\begin{table}[ht!]
\caption{Characteristics of MovieLens dataset}
\label{table:datasets1}
\centering
\begin{tabular}{|l |c |c |c| c| c| c|}
	\hline
	Dataset &   Users &  Movies & Ratings & Scale & Sparsity \\
	\hline
       Movielens 100K & $943$ & $1682$ & $100,000$ & $1$-$5$ & 94\% \\ \hline
       Movielens 1M  & $6040$ & $3952$ & $1,000,209$ & $1$-$5$ & 96\% \\
	\hline
\end{tabular}
\end{table}

\subsection{Experimental Settings}
We conduct experiments on two publicly available movie rating datasets: MovieLens 100K and MovieLens 1M \footnote{\url{https://grouplens.org/datasets/movielens/}}.
 The datasets are preprocessed, and users with less than $20$ observed ratings are removed.  The detailed characteristics of these datasets are reported in Table~\ref{table:datasets1}. Figure~\ref{fig:movielens} and \ref{fig:movielens1m} show the distribution of ratings in MovieLens 100K and MovieLens 1M datasets, respectively.

 We adopted the following two evaluation metrics most widely used to evaluate the recommendation performance of the methods, i.e.,  mean absolute error (MAE) and root mean square error (RMSE).

\begin{figure}[ht!]
    \centering
	\includegraphics[scale=0.6]{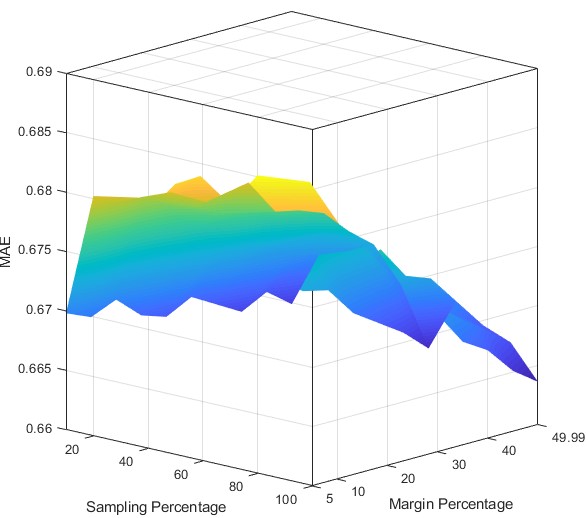}
	\caption{Influence of hyper-parameters $\tau_1$ and $s$ } 
    \label{fig:TauS}
\end{figure}

\begin{minipage}{\textwidth}
\begin{equation*}
   MAE  = \frac{\sum_{ij \in \Omega} | y_{ij} - \hat{y}_{ij}} | {|\Omega|},~~~
   RMSE = \sqrt{ \frac{\sum_{ij \in \Omega} (y_{ij} - \hat{y}_{ij} )^2 }{|\Omega|} },~~~~~~~~
\end{equation*}
\end{minipage}
\noindent
We consider SVD~\citep{koren2009matrix}, NMF~\citep{luo2014efficient}, SVD++~\citep{koren2008factorization} and Co-Clustering~\citep{george2005scalable} as our baseline algorithms for the performance evaluation. We have used the Surprise~\citep{Hug2020} library to run the baseline algorithms and tunning the hyperparameters after each stage of the augmentation process. For ST-MMMF, the regularization value $\lambda$ is tuned from the candidate set $\{10^\frac{i}{16}\}$, $\forall i \in \{1, 5, . . . , 40\}$. The value of sampling percentage $s$ in every iteration of ST-MMMF is searched in $\{10, 20, \dots, 100\}$. The parameter $\tau_1$ used to determine the high confidence region is tuned from $\{5, 10, 15, \dots, 45,49.99\}$ and the parameter $\tau_2$ is fixed to 10, defining it as the low confidence region and refined noise in each iteration. Figure~\ref{fig:TauS} shows the performance of ST-MMMF in terms of MAE on the Movielens 100K dataset. The MAE reported is an average of $50$ runs for every pair of $(\tau_1,s)$ for a fixed value of regularization parameter $\lambda$. We select $\tau_1 = 49.99$ and $s = 100$ corresponding to the smallest MAE for subsequent experiments on both datasets. Furthermore, we randomly selected $80\%$ of the observed ratings for training and the remaining $20\%$ as the test set. The prediction accuracy averaged over three runs is reported.

\subsection{Results and Analysis}

\begin{figure}[ht!]
\centering
\begin{subfigure}{.48\textwidth}
  \centering
  \includegraphics[width=1\textwidth, height = 5cm]{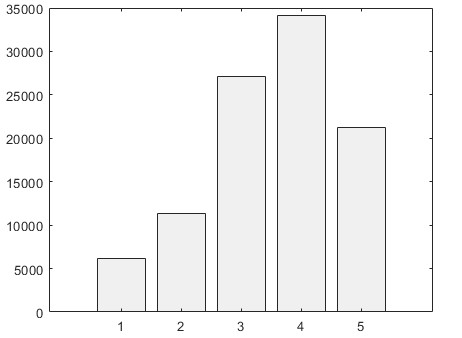}
	    \caption{MovieLens 100K }
	    \label{fig:movielens}    
\end{subfigure}%
  \hspace{.02\textwidth}
\begin{subfigure}{.48\textwidth}
  \centering
  \includegraphics[width=1\textwidth, height = 5cm]{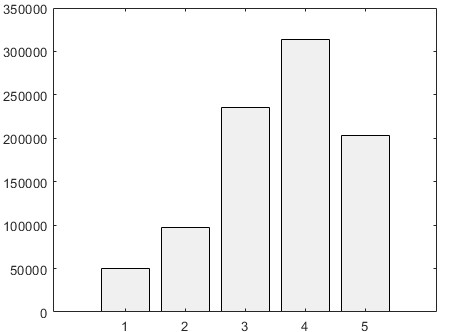}
    \caption{MovieLens 1M}
    	\label{fig:movielens1m}
\end{subfigure}
\caption{Rating Distribution}
\label{fig:test}
\end{figure}

We first analyze the effect of rating distributions on the prediction accuracy of each rating. It can be seen in ~\ref{fig:movielens} and \ref{fig:movielens1m} that the ratings exhibit imbalanced distribution, and the difference between the items rated 1 and 4 is significantly large. This would result in predictive bias where ratings with large samples may subsume ratings with a few samples~\citep{kumar2017collaborative}. Our augmentation process discussed in Section~\ref{PS1} takes care of  such imbalance rating distribution. 
To validate the proposed augmentation strategy, we have conducted some initial investigation using MMMF~\citep{rennie2005fast} over both datasets. 

\begin{table*}
    \caption{Effect of imputation on MovieLens 100K dataset. `*' indicates that the measure is not applicable. }
    \label{tab:ml100ImputationEffect}
    \begin{subtable}{1\linewidth}
      \centering
      \renewcommand*{\arraystretch}{1.1}
        \caption{Confusion matrix after the $1$st iteration. }
        \adjustbox{max width=\textwidth}{
         \begin{tabular}{|c|c|c|c|c|c|c|c|c|c|c|c|}
        \cline{1-12}
        &\multicolumn{11}{c|}{Predicted} \\\cline{2-12}
         & & 1 & 2 & 3 & 4 & 5 & $HR@0$ & $HR@1$ & $HR@2$ & $HR@3$ & $HR@4$\\  \cline{2-12}
        \multirow{13}{*}{\rotatebox{90}{Actual}}   &\multicolumn{11}{c|}{Training-set} \\ 
        \cline{2-12}
        & 1 & 2086 & 1865 & 864 & 70 & 3 & 0.4268 & 0.3815 & 0.1768 & 0.0143 & 0.0006  \\ \cline{2-12}
        & 2 & 160 & 3198 & 5296 & 432 & 10 & 0.3516 & 0.5998 & 0.0475 & 0.0011 & $*$ \\ \cline{2-12}
        & 3 & 9 & 225 & 16254 & 5173 & 55 & 0.7485 & 0.2486 & 0.0029 & $*$ & $*$ \\ \cline{2-12}
        & 4 & 0 & 10 & 2809 & 23892 & 628 & 0.8739 & 0.1257 & 0.0004 & 0 & $*$ \\ \cline{2-12}
        & 5 & 0 & 4 & 159 & 7030 & 9767 & 0.5759 & 0.4145 & 0.0094 & 0.0002 & 0\\ \cline{2-12}        
        &\multicolumn{11}{c|}{Test-set} \\ \cline{2-12} 
        & 1 & 231 & 248 & 505 & 230 & 8 & 0.1890 & 0.2029 & 0.4133 & 0.1882 & 0.0065\\ \cline{2-12}
        & 2 & 82 & 331 & 1263 & 583 & 15 & 0.1456 & 0.5915& 0.2564 & 0.0066 & $*$ \\ \cline{2-12}
        & 3 & 30 & 285 & 2843 & 2146 & 125 & 0.5237 & 0.4478 & 0.0286 & $*$ & $*$ \\ \cline{2-12}
        & 4 & 19 & 68 & 2033 & 4249 & 466 & 0.6217 & 0.3656 & 0.0099 & 0.0028 & $*$ \\ \cline{2-12}
        &  5 & 4 & 18 & 391 & 2692 & 1136 & 0.2679 & 0.6348 & 0.0922& 0.0042 & 0.0009\\ \cline{1-12}    
        \end{tabular} }
    \end{subtable} 
    
    \begin{subtable}{1\linewidth}
    \vspace{1cm}
      \centering
      \renewcommand*{\arraystretch}{1.1}
        \caption{Confusion matrix after the $50$th iteration.}
        \adjustbox{max width=\textwidth}{
         \begin{tabular}{|c|c|c|c|c|c|c|c|c|c|c|c|}
        \cline{1-12}
        &\multicolumn{11}{c|}{Predicted} \\\cline{2-12}
         & & 1 & 2 & 3 & 4 & 5 & $HR@0$ & $HR@1$ & $HR@2$ & $HR@3$ & $HR@4$\\  \cline{2-12}
        \multirow{13}{*}{\rotatebox{90}{Actual}}   &\multicolumn{11}{c|}{Training-set} \\ 
        \cline{2-12}
        & 1 & 26470 & 1522 & 1101 & 246 & 29 & 0.9013  & 0.0518 & 0.0375 & 0.0084 & 0.001 \\ \cline{2-12}
        & 2 & 296 & 39298 & 5050 & 881 & 64 & 0.8620  & 0.1173 & 0.0193 & 0.0014 & $*$ \\ \cline{2-12}
        & 3 & 124 & 570 & 61769 & 5653 & 258 & 0.9034  & 0.0910& 0.0056 & $*$ & $*$ \\ \cline{2-12}
        & 4 & 24 & 154 & 4519 & 66761 & 802 & 0.9239  & 0.0736 & 0.0021 & 0.0003 & $*$ \\ \cline{2-12}
        & 5 & 6 & 34 & 650 & 7708 & 56688 & 0.8710  & 0.1184 & 0.0100& 0.0005 & 0.0001 \\ \cline{2-12}
        
        &\multicolumn{11}{c|}{Test-set} \\ \cline{2-12} 
        & 1 & 243 & 237 & 505 & 229 & 8 & 0.1989  & 0.1939 & 0.4133 & 0.1874 & 0.0065 \\ \cline{2-12}
        & 2 & 79 & 341 & 1261 & 580 & 13 & 0.1500  & 0.5893 & 0.2551 & 0.0057 & $*$\\ \cline{2-12}
        & 3 & 28 & 261 & 2854 & 2164 & 122 & 0.5257  & 0.4467 & 0.0279 & $*$ & $*$ \\ \cline{2-12}
        & 4 & 18 & 67 & 1921 & 4348 & 481 & 0.6361  & 0.3514 & 0.0098 & 0.0026 & $*$ \\ \cline{2-12}
        & 5 & 4 & 17 & 383 & 2685 & 1152 & 0.2716  & 0.6331 & 0.0903 & 0.0040 & 0.0009  \\ \cline{1-12}         
        \end{tabular} }
    \end{subtable} 
\end{table*}

\begin{table*}
    \caption{Effect of imputation on MovieLens 1M dataset.  `*' indicates that the measure is not applicable. }
    \label{tab:ml1mImputationEffect}
    \begin{subtable}{1\linewidth}
      \centering
      \renewcommand*{\arraystretch}{1.1}
        \caption{Confusion matrix after the $1$st iteration.}
        \adjustbox{max width=\textwidth}{
         \begin{tabular}{|c|c|c|c|c|c|c|c|c|c|c|c|}
        \cline{1-12}
        &\multicolumn{11}{c|}{Predicted} \\\cline{2-12}
         & & 1 & 2 & 3 & 4 & 5 & $HR@0$ & $HR@1$ & $HR@2$ & $HR@3$ & $HR@4$\\  \cline{2-12}
        \multirow{13}{*}{\rotatebox{90}{Actual}}   &\multicolumn{11}{c|}{Training-set} \\ 
        \cline{2-12}
         & 1 & 27187 & 14822 & 2817 & 111 & 2 & 0.6050  & 0.3298 & 0.0627 & 0.0025 & 0 \\ \cline{2-12}
        & 2 & 1454 & 47408 & 35883 & 1296 & 4 & 0.5510  & 0.4336& 0.0151 & 0 & $*$ \\ \cline{2-12}
        & 3 & 40 & 3499 & 166131 & 39166 & 121 & 0.7950  & 0.2042 & 0.0008 & $*$ & $*$ \\ \cline{2-12}
        & 4 & 1 & 41 & 21005 & 251053 & 7076 & 0.8993  & 0.1006 & 0.0001 & 0 & $*$ \\ \cline{2-12}
        & 5 & 3 & 2 & 537 & 44048 & 136458 & 0.7537  & 0.2433 & 0.0030 & 0 & 0 \\ \cline{2-12}
        
        &\multicolumn{11}{c|}{Test-set} \\ \cline{2-12} 
        & 1 & 3530 & 3530 & 3473 & 643 & 59 & 0.3142  & 0.3142 & 0.3091 & 0.0572 & 0.0053 \\ \cline{2-12}
        & 2 & 2029 & 5314 & 9988 & 4106 & 75 & 0.2470  & 0.5586 & 0.1909 & 0.0035 & $*$ \\ \cline{2-12}
        & 3 & 1077 & 5133 & 23398 & 19980 & 2652 & 0.4479  &  0.4807 & 0.0714  & $*$ & $*$ \\ \cline{2-12}
        & 4 & 71 & 1909 & 17845 & 40173 & 9797 & 0.5756  & 0.3960 & 0.0274 & 0.0010 & $*$ \\ \cline{2-12}
        & 5 & 32 & 89 & 4966 & 22351 & 17824 & 0.3938  & 0.4938 & 0.1097 & 0.0020 & 0.0007 \\ \cline{1-12}       
        \end{tabular} }
    \end{subtable} 

    \begin{subtable}{1\linewidth}
     \vspace{1cm}
      \centering
      \renewcommand*{\arraystretch}{1.1}
        \caption{Confusion matrix after the $50$th iteration.}
        \adjustbox{max width=\textwidth}{
         \begin{tabular}{|c|c|c|c|c|c|c|c|c|c|c|c|}
        \cline{1-12}
        &\multicolumn{11}{c|}{Predicted} \\\cline{2-12}
         & & 1 & 2 & 3 & 4 & 5 & $HR@0$ & $HR@1$ & $HR@2$ & $HR@3$ & $HR@4$\\  \cline{2-12}
        \multirow{13}{*}{\rotatebox{90}{Actual}}   &\multicolumn{11}{c|}{Training-set} \\ 
        \cline{2-12}
         & 1 & 84016 & 13840 & 3267 & 235 & 39 & 0.8286  & 0.1365 & 0.0322 & 0.0023 & 0.0004  \\ \cline{2-12}
        & 2 & 2236 & 100004 & 35825 & 1727 & 48 & 0.7151  & 0.2722 & 0.0123 & 0.0003 & $*$  \\ \cline{2-12}
        & 3 & 281 & 4684 & 209436 & 39915 & 472 & 0.8220   & 0.1750 & 0.0030 & $*$ & $*$\\ \cline{2-12}
        & 4 & 68 & 323 & 24209 & 287691 & 8165 & 0.8978  & 0.1010 & 0.0010 & 0.0002 & $*$  \\ \cline{2-12}
        & 5 & 17 & 74 & 949 & 44973 & 182672 & 0.7988   & 0.1967 & 0.0041 & 0.0003 & 0.0001 \\ \cline{2-12}
        
        &\multicolumn{11}{c|}{Test-set} \\ \cline{2-12} 
        & 1 & 3587 & 3502 & 3450 & 640 & 56 & 0.3193   & 0.3117 & 0.3071 & 0.0570 & 0.005 \\ \cline{2-12}
        & 2 & 1992 & 5418 & 9960 & 4083 & 59 & 0.2519   & 0.5556 & 0.1898 & 0.0027 & $*$ \\ \cline{2-12}
        & 3 & 1057 & 5082 & 23525 & 19956 & 2620 & 0.4503  & 0.4793 & 0.0704 & $*$ & $*$  \\ \cline{2-12}
        & 4 & 40 & 1854 & 17799 & 40370 & 9732 & 0.5784  & 0.3945 & 0.0266 & 0.0006 & $*$  \\ \cline{2-12}
        & 5 & 21 & 75 & 4886 & 22285 & 17996 & 0.3976   & 0.4923 & 0.1079 & 0.0017 & 0.0005 \\ \cline{1-12}        
        \end{tabular} }
    \end{subtable} 
\end{table*}

Table~\ref{tab:ml100ImputationEffect} and~\ref{tab:ml1mImputationEffect} show the effect of augmentation on individual rating prediction using MovieLens 100K and  MovieLens 1M datasets, respectively. We also calculated label-wise statistics to measure the number of times, on average, the actual and predicted rating labels differ with $\pm K$ distance. For simplicity, we denoted this as $HR@K$, where $HR@0$ for any $r$th rating label is the fraction of \textit{hits} to the $r$th rating label in the prediction. As discussed, this statistic is more useful in discrete ordinal rating prediction. It is always better to predict `2' than `5' if the actual label is `1', unlike the multi-class classification setting where all mistakes are equal. The proposed augmentation strategy leads to a more balanced dataset and improves individual label prediction accuracy.

\begin{figure}[ht!]
     \centering
     \begin{subfigure}[b]{0.46\textwidth}
         \centering
         \adjustbox{max width=\textwidth}{
         \includegraphics[scale=1]{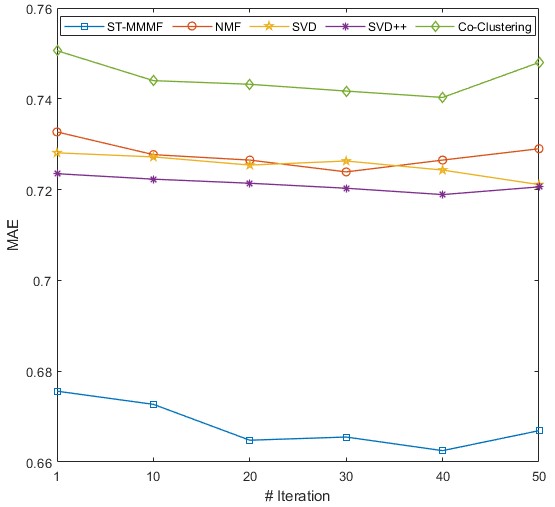}
         }
         \caption{MAE for MovieLens 100K}
         \label{fig:ML100K_MAE}
     \end{subfigure}%
     \begin{subfigure}[b]{0.46\textwidth}
         \centering
         \adjustbox{max width=\textwidth}{
         \includegraphics[scale=1]{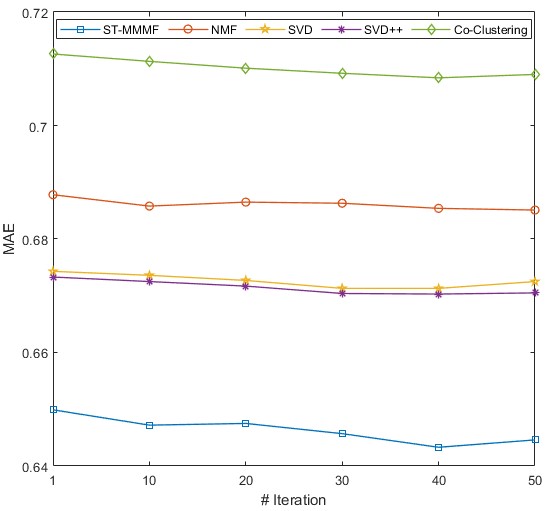}
         }
         \caption{MAE for MovieLens 1M}
         \label{fig:ML1M_MAE}
     \end{subfigure}\hfill
     \begin{subfigure}[b]{0.46\textwidth}
         \centering
         \adjustbox{max width=\textwidth}{
         \includegraphics[scale=1]{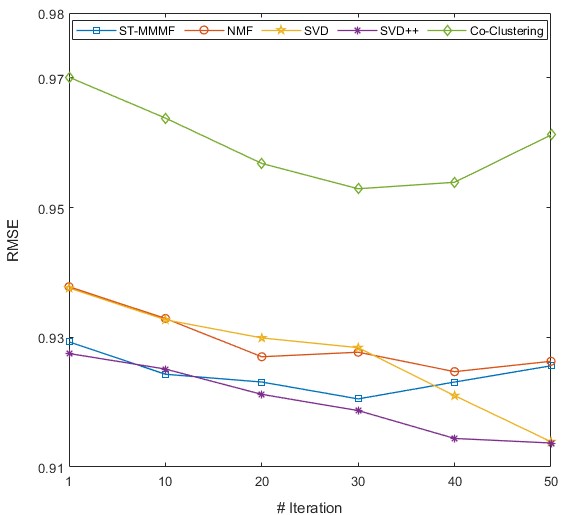}
         }
         \caption{RMSE for MovieLens 100K}
         \label{fig:ML100K_RMSE}
     \end{subfigure}%
     \begin{subfigure}[b]{0.46\textwidth}
         \centering
         \adjustbox{max width=\textwidth}{
         \includegraphics[scale=1]{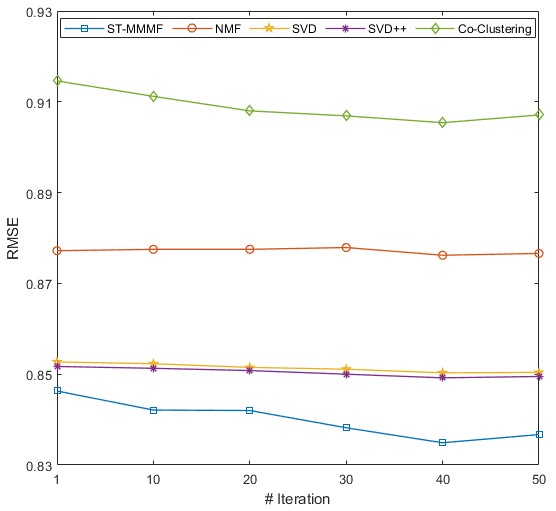}
         }
         \caption{RMSE for MovieLens 1M}
         \label{fig:ML1M_RMSE}
     \end{subfigure}
     \caption{ Performance of baseline algorithms on MovieLens 100K dataset}
        \label{fig:movielensBasline}
\end{figure}

\noindent We next evaluated the performance of each baseline algorithm with the augmented training set. After each round of the augmentation process, the baseline models are retrained, and their performance over the test set is recorded. Figure~\ref{fig:movielensBasline} reports the results related to this experiment. It can be seen that the process of data augmentation brings substantial improvements to the performance of each baseline algorithm. Furthermore, it can also be seen that the performance of several baseline algorithms starts decreasing after a few rounds of augmentation. The reason is that test sets are created before the augmentation process begins. Hence, the proportion of the rating label in the test set nearly follows the same initial rating distribution. After several rounds of the augmentation process, the rating distribution in the training set becomes more balanced. In contrast, the rating distribution in the test set remains unchanged, leading to decreased performance over the test set. This result prompts a new line of future research of deciding the maximum tolerable percentage of ratings that can be imputed to the rating matrix.

\begin{table}[ht!]
\caption{Effect of data augmentation}
\adjustbox{max width=\textwidth}{
\begin{tabular}{|p{0.15\textwidth} |p{0.15\textwidth}|p{0.18\textwidth}|p{0.25\textwidth}|p{0.14\textwidth}|p{0.25\textwidth}|}
\hline
\textbf{\# Iteration} & \textbf{\# Observed entries} &  \textbf{\# Unobserved entries} & \textbf{\# Unobserved entries  predicted with high-confidence}

 & \textbf{\# Ratings augmented }& \textbf{\# Overlap between the ratings predicted with high confidence} \\ \hline
1  & 80000  & 1506126 & 395851 & 5000 & NA     \\ \hline
2  & 85000  & 1501126 & 485458 & 5000 & 374637 \\ \hline
3  & 90000  & 1496126 & 538751 & 5000 & 467698 \\ \hline
4  & 95000  & 1491126 & 565265 & 5000 & 519462 \\ \hline
5  & 100000 & 1486126 & 604835 & 5000 & 552333 \\ \hline

\end{tabular}
\label{tab:properties}
}
\end{table}

\noindent We also experiment to validate that the proposed rating augmentation process satisfies certain desirable properties such as \textit{monotonicity} and \textit{invariance}. By \textit{monotonicity}, we mean that the proposed augmentation process exhibits significant overlap among the ratings predicted with high confidence across iterations. In other words, The entries that satisfy Equation~(\ref{eq:augment}) in an iteration continue to be of high confidence in subsequent iterations. The augmentation process is said to be \textit{decision invariant} if the percentage of ratings predicted with high confidence increases in each subsequent iteration. We have reported the output of the first five runs of ST-MMMF on MovieLens 100K in Table~\ref{tab:properties}. It can be seen from the table that our proposed approach satisfies both the crucial properties of the augmentation process. We have observed similar performance in the refinement process.

\section{Conclusions and Future Work}\label{CFW}
This paper presented a novel self-training-based semi-supervised
approach, ST-MMMF,  for rating augmentation and refinement that uses maximum-margin matrix factorization (MMMF) as the base learner. The proposed rating augmentation and refinement process exploits the geometrical interpretation of MMMF. The proposed approach is an iterative scheme that uses some of the ratings predicted with high confidence in every iteration to augment the training set. The ratings predicted with low confidence are removed from the training set in the refinement phase. Extensive experimental studies performed over the two highly imbalanced real-world rating datasets corroborate our claim that the proposed approach significantly reduces the prediction bias of MMMF toward rating labels of high samples. Furthermore, we also evaluated the performance of several state-of-the-art algorithms to validate that the proposed rating augmentation strategy is likely to
reduce or alleviate the data sparsity problem. 

There are numerous avenues for future research, addressing several issues with the proposed approach and possible modeling extension to different application areas. It is interesting to see whether ST-MMMF can be further improved by considering side information from other auxiliary domains, thereby enabling a more robust and meaningful confidence association with the predicted ratings. This line of research is more applicable to scenarios such as route recommendations that require the incorporation of multi-modal transport and locations of facilities/services~\citep{zhang2011multimodal, liao2013incorporating}. Extension of the proposed approach to diverse applications, especially in scenarios where abundant information is unavailable for the modeling, such as course~\citep{parameswaran2011recommendation} and tourism destination recommendations~\citep{lucas2013hybrid}, presents a compelling direction for further exploration. Furthermore, a significant issue worth exploring involves applying the proposed approach in the problem area where the observed entries do not have any predefined order. A comprehensive study in this direction is required where the objective function of the underlying maximum margin matrix factorization algorithm needs to be carefully modified to align with the principles of multi-class classification problems while preserving the maximum margin property. We plan to investigate these aspects in the future.

\section*{Acknowledgements}

Vikas Kumar is supported by the Start-up Research Grant (SRG), Science and Engineering Research Board, under grant number SRG/2021/001931, UGC-BSR Start-up Grant, UGC under grant number F.30-547/2021(BSR) and the Faculty Research Programme Grant, University of Delhi, under grant number IoE/2021/12/FRP. 

\bibliographystyle{model5-names}
\biboptions{authoryear}
\bibliography{reference}

\end{document}